\documentclass[letterpaper]{mn2e}
\usepackage{epsfig,natbib2,natbibmnfix}

\newcommand{\be}{\begin{equation}}
\newcommand{\ee}{\end{equation}}

\newcommand{\apj}{ApJ}

\newcommand{\mnras}{MNRAS}
\newcommand{\aap}{A\&A}

\newcommand{\apjl}{ApJL}

\newcommand{\nat}{Nature}

\def\ltsima{$\; \buildrel < \over \sim \;$}
\def\simlt{\lower.5ex\hbox{\ltsima}}
\def\gtsima{$\; \buildrel > \over \sim \;$}
\def\simgt{\lower.5ex\hbox{\gtsima}}
\newcommand\sgra{Sgr~A$^*$}
\newcommand\medd{\dot{M}_{\rm Edd}}
\newcommand\mcapt{\dot{M}_{\rm capt}}

\newcommand\mdot{\dot{m}}

\def\msun{{\,{\rm M}_\odot}}
\newcommand\mbh{{\,{\rm M}_{\rm bh}}}

\def\del#1{{}}

\title{X-rays from cusps of compact remnants near galactic centres}

\author[S.~Nayakshin \& R. Sunyaev]
{\parbox{18cm}{Sergei Nayakshin$^{1}$ and Rashid Sunyaev$^{2}$}\vspace{0.3cm}\\
$^1$Dept. of Physics \& Astronomy, University of Leicester, Leicester, LE1 7RH,
UK\\
$^2$Max-Planck-Institut f\"{u}r Astrophysik, Karl-Schwarzschild-Stra\ss{}e 1,
85741 Garching bei M\"{u}nchen, Germany}

\begin{document}

\maketitle

\begin{abstract}
Compact remnants -- stellar mass black holes and neutron stars formed in the
inner few parsec of galactic centres are predicted to sink into the central
parsec due to dynamical friction on low mass stars, forming a high
concentration cusp \citep{Morris93}. Same physical region may also contain
very high density molecular clouds and accretion discs that are needed to fuel
SMBH activity.  Here we estimate gas capture rates onto the cusp of stellar
remnants, and the resulting X-ray luminosity, as a function of the accretion
disc mass. At low disc masses, most compact objects are too dim to be
observable, whereas in the high disc case most of them are accreting at their
Eddington rates. We find that for low accretion disc masses, compact remnant
cusps may be more luminous than the central SMBHs. This ``diffuse'' emission
may be of importance for local moderately bright AGN, especially Low
Luminosity AGN. We also briefly discuss how this expected emission can be used
to put constraints on the black hole cusp near our Galactic Centre.
\end{abstract}

\renewcommand{\thefootnote}{\fnsymbol{footnote}}
\footnotetext[1]{E-mail: {\tt Sergei.Nayakshin at astro.le.ac.uk}}

\section{Introduction}
\label{intro}

Super-massive black holes are predicted to be surrounded by clusters of
compact stellar remnants, i.e., stellar mass black holes and neutron
stars. Dynamical friction \citep{Chandrasekhar43} on less massive background
stars transfers these remnants closer to the SMBH \citep{Morris93}. For the
case of our Galactic Centre, stellar mass black holes can be collected into
the central parsec from a sphere with radius about 4 parsec
\citep{Morris93}. \cite{Miralda00} updated and quantified \cite{Morris93}
estimates and showed that black holes will force most of low mass stars out of
the central region with size $R_{\rm bh} \approx 0.7$ parsec around \sgra. In
their model, black holes form a central cluster of this size and number about
$20,000$. The total mass of these black holes is around $5\%$ of \sgra's mass
\citep{Schoedel02,Ghez03a}. \cite{Freitag06} have recently modelled the
dynamical mass segregation in the central few parsec of our Galaxy with state
of the art Monte Carlo methods, largely confirming results of \cite{Morris93}
and \cite{Miralda00}. In addition, it was shown that the central parsec
contains neutron stars in numbers comparable to that of the black holes.

Recent observations of young massive stars near \sgra\ suggest that these
stars may also be born in situ with an unusually top-heavy mass function
\citep{Paumard06,NS05,Nayakshin06a}. The remnants of the young stars born in
situ will contribute to the cusp of black holes and neutron stars formed by the
dynamical friction.

\cite{Morris93} discussed the possibility that compact remnants will be
accreting gas from molecular clouds that may be unusually dense near our GC.
In this paper we set out to explore this idea in more detail and also
speculate on the implications for observations of AGN in general. Our model
includes a realistic form for the radiative efficiency of these sources
motivated by the observations of radiatively inefficient accretion flows in
X-ray binaries. We integrate over the black hole velocity distribution to
obtain a time-average estimate of total X-ray luminosity of the cluster as a
function of the mass of the embedded gaseous accretion disc. 

We find that at low disc masses, most of the stellar remnants accrete at
highly sub-Eddington radiatively inefficient rates, and are quite
dim. However, there are always few that almost co-move with the gas in the
disc. These objects achieve relatively large accretion rates. Since the SMBH
itself becomes radiatively inefficient at low disc masses, our model predicts
that clusters of stellar remnants may actually outshine the SMBH in this Low
Luminosity AGN (LLAGN) regime.  At the high luminosity, high disc mass regime,
the super-massive black hole should dominate bolometrically. Stellar remnants
may however be important in certain wavelengths.

\section{The model}

\subsection{Bondi-Hoyle gas capture rates}

We shall first make order of magnitude estimates to motivate a more detailed
treatment.  A stellar mass black hole of mass $\mbh$, ploughing through a gas
cloud, is capturing gas at the \cite{Bondi44} accretion rate:
\begin{equation}
\mcapt \sim 4 \pi \rho \frac{(G \mbh)^2}{(\Delta v^2 + c_s^2)^{3/2}}\;,
\end{equation}
where $\rho$, $c_s$ and $\Delta v$ are the ambient gas density, sound speed
and the relative velocity between the black hole and the gas, respectively.

Consider first the situation in the central parsec of our Galaxy.  Stellar
velocity dispersion, $\sigma$, is of the order of a few hundred km/sec there.
The number of black holes, $\Delta N_{\rm bh}$, moving with respect to gas
with a relative velocity smaller than $\Delta v$ is of the order of $\Delta
N_{\rm bh} \sim N_{\rm bh} (\Delta v/v_{\rm esc})^3 f_v$, where $ N_{\rm bh}
\gg 1$ is the total number of stellar mass black holes, $v_{\rm esc}\sim
\sigma$ is the escape velocity, and $f_v \ll 1$ is the volume filling factor
of cold gas in the region. For an estimate, take $f_v = 0.01$ and $N_{\rm bh}
= 10^4$. In that case, $\Delta N_{\rm bh} \sim 1$ for $\Delta v \simgt 50$
km/sec. There thus should be few stellar mass black holes travelling through
the gas at $\Delta v \sim 50$ km/sec. Neglecting the gas sound speed in
equation 1, the Bondi-Hoyle capture rate is
\begin{equation}
\mcapt \sim 4.8 \times 10^{-15} \; \msun \hbox{year}^{-1} \; \frac{n_0
\mu}{\Delta v_{50}^3}\;,
\label{bh}
\end{equation}
where $n_0$ is the gas density in Hydrogen atoms per cm$^3$, $\mu$ is mean
molecular weight in units of proton mass, and $\Delta v_{50} \equiv \Delta
v/(50)$ km/sec.

This accretion rate is quite small compared to the Eddington limit of $\medd
\approx 2\times 10^{-7} (\mbh/10\msun) \msun$ year$^{-1} $ if $n_0
=1$. However, gas density can be much higher. The hot X-ray emitting gas in
the inner parsec of our Galaxy has density of $n \sim 10-100$ cm$^{-3}$
\citep{Baganoff03a}. If we assume cool gas clouds to be in pressure
equilibrium with the hot gas, then the cool gas density could be as high as
$10^5 - 10^6$ cm$^{-3}$. Therefore we could in principle expect several
stellar mass black holes accreting gas at rates corresponding to $\sim
10^{-4}-10^{-2}$ of their Eddington rates.  Since \sgra\ produces only $\sim
10^{33}$ erg/sec in X-rays \citep{Baganoff03a}, the brighter of the accreting
black holes can actually outshine \sgra\ in X-rays.

In the case of a reasonably bright quasar or an AGN accreting through a parsec
scale accretion disc, we can estimate the gas density in the disc in the
following way. In order to be stable against self-gravitational collapse, the
disc's \cite{Toomre64} $Q$-parameter should be smaller than unity
\citep[e.g.,][]{Gammie01,Goodman03,Thompson05,Nayakshin06a}. The gas density
in the disc satisfies
\begin{equation}
\rho = Q^{-1} \rho_{\rm crit} \equiv Q^{-1} \frac{M_{\rm SMBH}}{\sqrt{2} \pi
R^3}\;,
\label{rho}
\end{equation}
where $\rho_{\rm crit}$ is the critical density at which the disk is
marginally self-gravitating. Numerically,
\begin{equation}
n = \frac{\rho}{\mu m_p} = 2.8 \times 10^{12} \hbox{cm}^{-3} Q^{-1}\,
\frac{M_8}{\mu R_{0.1}^3}\;,
\end{equation}
where $M_8 = 10^{-8} M_{\rm SMBH}/\msun$, $R_{0.1} = R/0.1$ parsec and $m_p$
is proton's mass. For such large gas densities, even black holes with very
large differential velocity $\Delta v$, e.g., $\Delta v \simgt 1000$ km/s,
would accrete at their super-Eddington rates:
\begin{equation}
\mcapt \sim 100 \; \medd \; \frac{M_8}{q R_{0.1}^3 \Delta v_{1000}^3}\;.
\end{equation}
In practice, this implies that {\em every} black hole crossing the AGN
accretion disc captures the material at or above its Eddington accretion rate.

\subsection{Maximum accretion rate}

A black hole co-moving with the gas, e.g. with $\Delta v = 0$, is predicted to
accrete at a very large accretion rate by the equation 1. However, in reality
gas must be in a Keplerian differential rotation about the SMBH. Thus, even if
gas exactly coincident with the stellar mass black hole has zero relative
velocity, the neighbouring regions will have $\Delta v \ne 0$. In such
situation, the sphere of influence of the stellar mass black hole is the
Hill's radius $r_H = R (\mbh/3M_{\rm SMBH})^{1/3}$ \citep{Lissauer87}, and the
maximum accretion rate is of order of
\begin{equation}
\dot M_{\rm H} = 4 \pi r_H^2 \rho c_s\;.
\end{equation}
We thus require 
\begin{equation}
\dot M_{\rm bh} = \hbox{min}\;\left[\dot M_{\rm capt}, \dot M_{\rm
H}\right]\;.
\end{equation}

\subsection{Expected accretion luminosity}

As is well known, standard accretion discs have a very high radiative
efficiency: $\epsilon_0 = 0.1 c^2$ \citep{Shakura73}. However, at low
dimensionless accretion rates, $\mdot = \dot M/\medd \ll \mdot_{\rm crit}$,
radiative efficiency is predicted to be much smaller since electrons in the
flow may be cooler than protons \citep{Narayan95} and since gas outflows may
be important \citep{Blandford99}. For definitiveness, we follow the
\cite{Narayan95} model, in which $\epsilon = \epsilon_0 (\mdot/\mdot_{\rm
crit})$ for $\mdot < \mdot_{\rm crit}$, and $\mdot_{\rm crit}=0.01$.  This
model seems to be compatible with X-ray observations
\citep[e.g.,][]{Esin97}. Joining the two accretion regimes smoothly, and
assuming that roughly 10\% of the bolometric luminosity comes out in the X-ray
band, we write
\begin{equation}
\epsilon(\mdot) = \epsilon_{0, \rm x} \frac{\mdot}{\mdot + \mdot_{\rm
crit}}\;,
\label{eff}
\end{equation}
where $\epsilon_{0, \rm x} = 0.01$. For accreting neutron stars, the
radiative efficiency is assumed to be always large and equal to $\epsilon_{0,
\rm x}$ due to the presence of hard surface.

\subsection{Accretion and emission from the black hole cluster}

Black holes in collisionless cusps around SMBHs are expected to have the
isotropic \cite{Bahcall76} space-velocity distribution. This distribution is a
good approximation to the recent detailed Monte-Carlo simulations
\citep{Freitag06}, in which the radial black hole density distribution found
to follow the law $\rho_{bh}(R) \propto R^{-7/4}$. To keep our analysis
analytically simple, we approximate this law to $\rho_{bh}(R) \propto
R^{-3/2}$, in which case the black hole space and velocity distribution is
particularly simple \citep[e.g.,][]{Miralda00}:
\begin{equation}
f({\bf R}, {\bf v}) = \cases{C_0, \; \hbox{if} \quad v \le v_{esc},\cr 0, \quad
\hbox{otherwise.}}
\end{equation}
where we also assume $R \le R_{\rm bh}$, the radius of the black hole cluster
around the SMBH. The constant $C_0$ is set by requiring $N_{\rm bh} = \int
d{\bf R} \int d{\bf v} f({\bf R}, {\bf v})$. Using this velocity distribution,
we calculate the total accretion rate onto the stellar mass black holes,
assuming that gas moves on circular Keplerian orbits in a thin disk of half
thickness $H\ll R$.
\begin{equation}
\dot M_{\rm tot} = \int d{\bf R} \rho({\bf R}) \int d{\bf v} f({\bf R}, {\bf
v}) \frac{\dot M_{\rm bh}}{\rho({\bf R})}\;.
\label{mtot}
\end{equation}
Note that, except for very high accretion rates, $\dot M_{\rm bh}= \dot M_{\rm
capt}\propto \rho$, and hence the expression under the velocity integral does
not depend on $\rho({\bf R})$. We make one further approximation, writing
$v_{\rm esc}(R) \approx v_{\rm esc}(R_{\rm bh})$. This is justified since most
emission arises from the edge of the black hole cluster, i.e. $R\sim R_{\rm
bh}$. This is so because that region contains most of black holes and they
also have the smallest $\Delta v$, as $v_{\rm esc}$ is smaller than in the
inner regions. We thus have,
\begin{equation}
\dot M_{\rm tot} \approx M_{\rm disc} \int_{v \le v_{\rm esc}(R_{\rm bh})}
d{\bf v} C_0 \frac{\dot M_{\rm bh}}{\rho(R_{\rm bh})}\;.
\label{mtotap}
\end{equation}
In this simple model, the total accretion rate onto stellar mass black holes
scales linearly with the disc mass. The total luminosity of the black hole
cluster is given by a similar equation to \ref{mtot} but with the additional
$\epsilon(\mdot)$ (equation \ref{eff}) included under the sign of the
integral.

A more detailed treatment of this problem is underway (Deegan \& Nayakshin, in
preparation). In this Monte Carlo--like approach, we follow trajectories of
accreting objects explicitly, and treat both capture of gas and ``small''
scale disc accretion in a time-dependent manner. The results are consistent
with those obtained in this paper but show a large time-variability, as
expected.

\subsection{Model of disk accretion for the SMBH}

 In the absence of a reliable theory which would connect accretion disk
properties at parsec scales to the inner regions of accretion flows on SMBH at
both low and high mass accretion rate regimes
\citep[e.g.,][]{Goodman03,Cuadra05}, we model the accretion onto the SMBH in
the following way. We estimate the viscous time scale for accretion through
the disk as $t_{\rm visc} = \alpha^{-1} (H/R)^{-2} \Omega^{-1}$
\citep{Shakura73}. The accretion rate onto the SMBH is then
\begin{equation}
\dot M_{\rm SMBH} \sim \frac{M_{\rm disc}}{t_{\rm visc}}= \alpha
\frac{H^2}{R^2} \Omega M_{\rm disc}\;,
\end{equation}
The disc height scale $H$ is calculated assuming the hydrostatic balance as
long as the disc is non-self-gravitating (i.e., relatively low-mass).  For
simplicity we assume disk temperature is equal to 100 Kelvin. The results
depend weakly on this value as $H/R \propto T^{1/2}$ in this regime. For more
massive self-gravitating disks \citep[see,
e.g.,][]{Paczynski78,Gammie01,Goodman03,NC05}, we assume $Q\sim 1$, which
translates into $H/R \sim M_{\rm disc}/M_{\rm SMBH}$.  The SMBH luminosity is
calculated in the same way as for the stellar mass black holes, i.e. $L_{\rm
SMBH} = \epsilon(\mdot) \dot M_{\rm SMBH}$, where $\epsilon(\mdot)$ is given
by equation \ref{eff}.

\section{Results and Observational Implications}\label{sec:results}

\begin{figure*}
\begin{minipage}[b]{.49\textwidth}
\centerline{\psfig{file=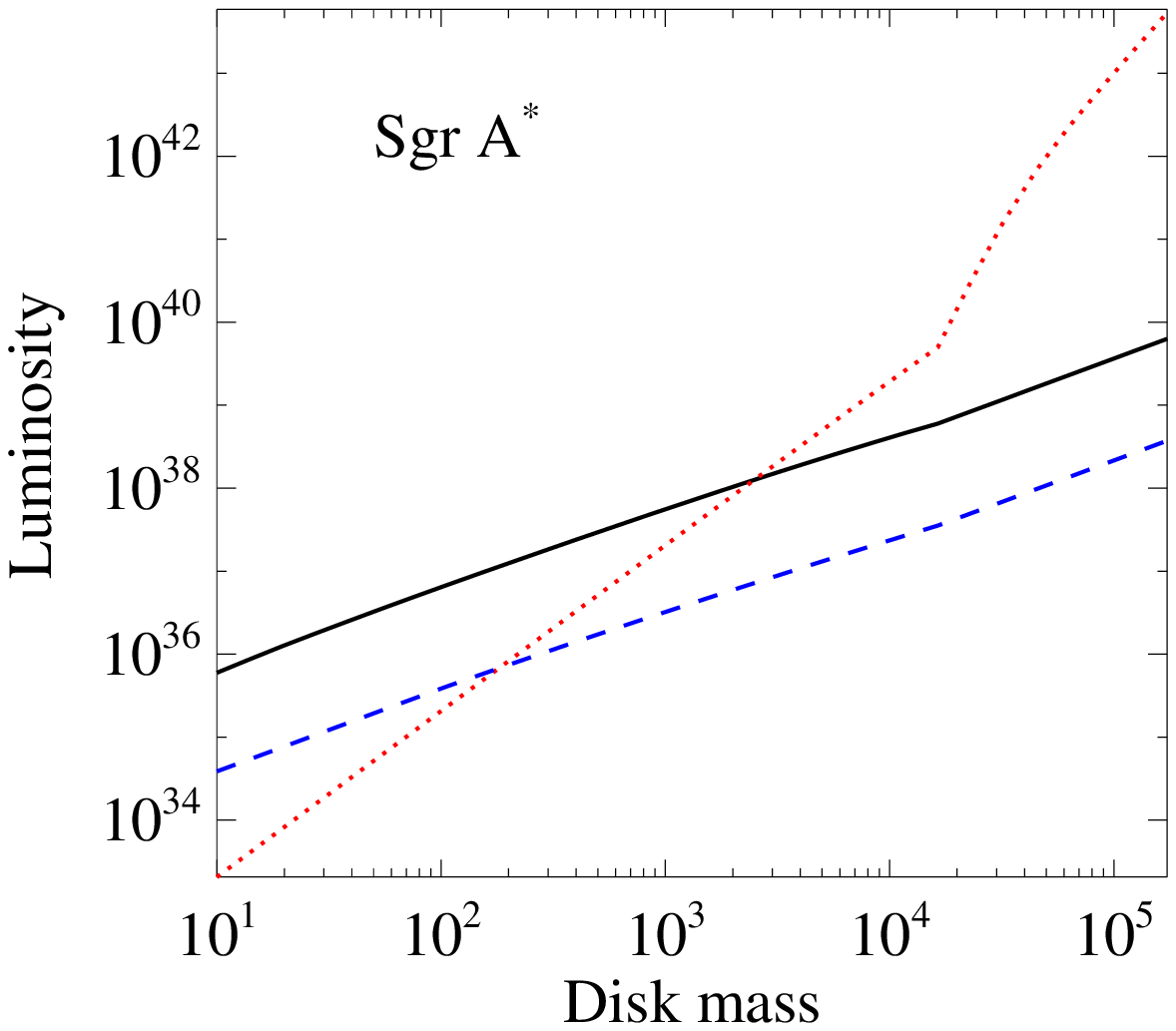,width=.99\textwidth,angle=0}}
\end{minipage}
\begin{minipage}[b]{.49\textwidth}
\centerline{\psfig{file=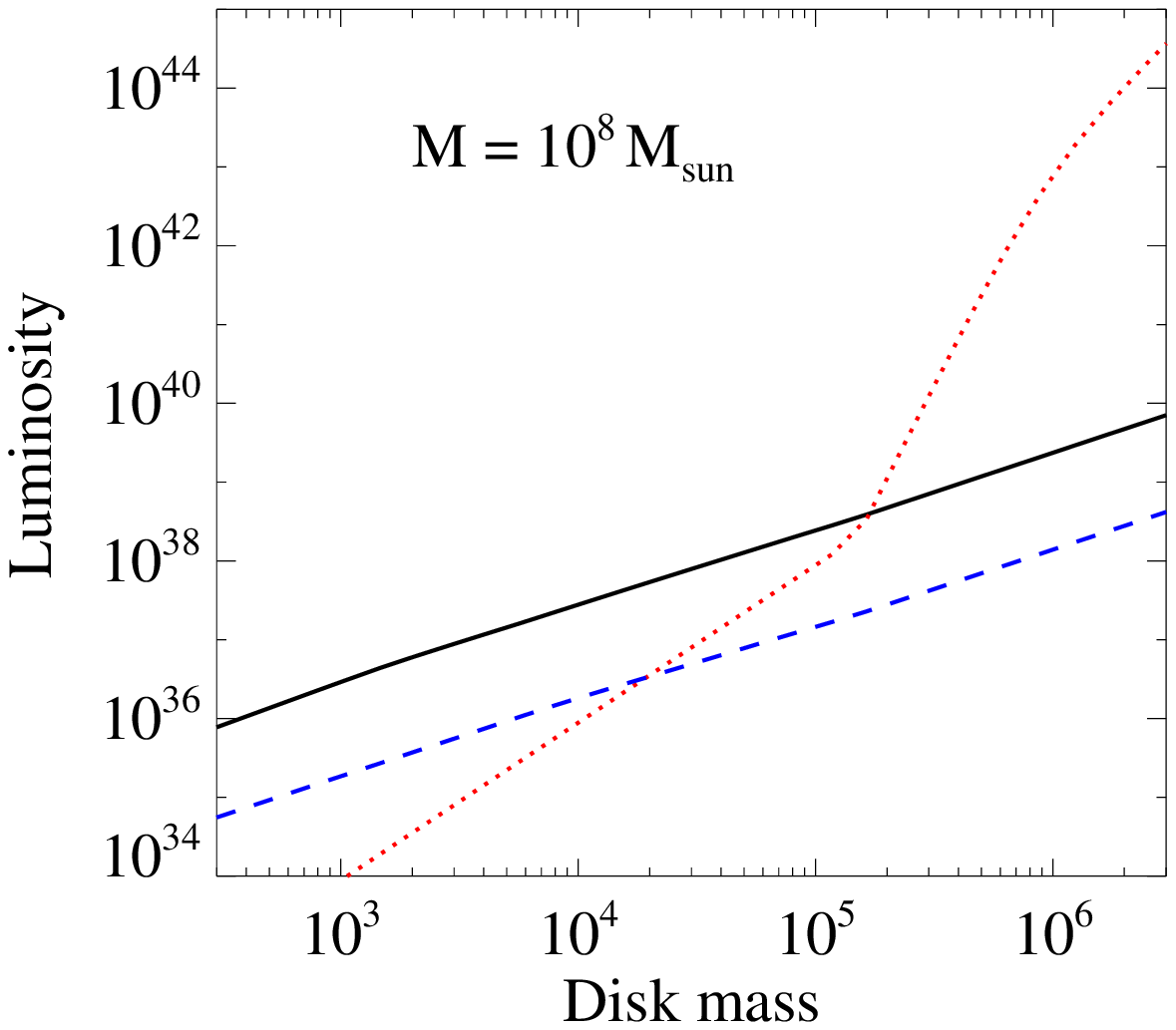,width=.99\textwidth,angle=0}}
\end{minipage}
\caption{The X-ray luminosity (in erg s$^{-1}$) of the different accreting
components versus the accretion disc mass (in units of $\msun$). In
particular, the solid, dashed and dotted curves show the luminosities of the
black hole cusp, the neutron star cusp, and the SMBH, respectively. The left
panel is for the case of \sgra, and the right one for a SMBH with $\mbh = 10^8
\msun$. Note that at low disc masses (low SMBH luminosities), the cusps of
compact remnants are brighter than the central massive black hole.}
\label{fig:fig1}
\end{figure*}

With this model we now calculate two examples. The first one is based on the
up to date models of the central parsec of our Galaxy
\citep[e.g.,]{Freitag06}, and the second is for the $M=10^8 \msun$ SMBH case.
For both of these we assumed the total mass of the stellar mass black holes to
be 5\% of the mass of the central SMBH. For the GC, this yields $N_{\rm bh} =
24,000$, whereas $N_{\rm bh} = 720,000$ for the more massive galactic
nucleus. The black hole cluster is assumed to extend to $R_{\rm bh}=0.7$ pc
for \sgra\ case \citep{Miralda00}, and to $R_{\rm bh}=2$ pc for the other
case. We also calculate the expected X-ray emission from neutron stars in the
cluster, taken to be as numerous as the stellar mass black holes
\citep{Freitag06} but have the mass of 1.4 $\msun$. Note that in this paper we
consider neutron stars to be magnetic field free. If their fields are strong
enough, the ``propeller effect'' may reduce the X-ray luminosity of neutron
stars \citep{MenouEtal99}.

Figure \ref{fig:fig1} shows the X-ray luminosities of: the SMBH (dotted), the
cluster of stellar mass black holes (solid), and the neutron star cluster
(dashed) as a function of the accretion disc mass.  The left panel shows the
GC case, whereas the right panel shows the $M_{\rm SMBH}=10^8 \msun$ SMBH
case.  The main prediction of Figure \ref{fig:fig1} is that at high disc
masses the black hole cluster emission is negligible compared to that of the
SMBH, but at smaller disc masses the situation is reversed. This is understood
as following. At the highest disc masses, all of the components considered
accrete at rates approaching their respective Eddington limits. As the SMBH is
more massive than the cusps we model, it is brighter than the stellar mass
compact remnants. At very low disc masses, on the contrary, the SMBH accretes
in the radiatively inefficient regime, and it is quite dim. At the same time,
the accretion rates onto the stellar mass black holes span a wide range
according to a range in the relative velocities $\Delta v$. Though small in
numbers, black holes with small relative velocities dominate the radiative
output of the cluster since the radiative efficiency for them is largest (see
equation \ref{eff}). We now discuss the observational implications of our
results in greater detail.


\subsection{``Fake X-ray binaries'' in \sgra}

It is only in the centre of our own Galaxy that we have observational
constraints on the distribution of gas and X-ray sources in the inner parsec.
The dominant reservoir of cold gas in the inner parsec of our GC is the
so-called minisprial \citep{Scoville03,Paumard04}. This morphologically
complicated feature covers a good fraction of the inner parsec in
projection. It is believed to be mildly geometrically thick, and have
temperature and density of order of $T\sim 6\times10^3$ K and $n_0 \simgt
10^4$, respectively. The high temperature is maintained by photo-ionisation
via radiation of nearby luminous massive stars. The mass of the mini-spiral
was estimated at $\sim 50 \msun$ \citep{Paumard04}, although observational
uncertainties are large, and the real mass could be as high a factor of 10
larger (T. Paumard, private communication).

With $M_{\rm disc} = 50 \msun$, the left panel of Figure 1 predicts that the
X-ray luminosity of \sgra\ should be $\sim$few$\times 10^{34}$ erg/s. This is
about a factor of ten larger than actually observed \citep{Baganoff03a}. Given
many approximations and uncertainties that are used in our simple model, the
disagreement is not surprising, especially since the fraction of radiation
power coming out in X-rays is exceptionally low for \sgra\ (which is not taken
into account in our model). 

Figure 1 further shows that the population of 24,000 black holes would emit
around $10^{36}$ erg/s in X-rays.  Most of this luminosity comes from a few to
a dozen of X-ray sources with X-ray luminosities ranging between $10^{33}$ to
$10^{35}$ erg/s. It thus seems possible that some of the {\em unusual} X-ray
sources in the inner parsec detected by \cite{Muno05} are the isolated
accreting black holes or neutron stars masquerading themselves as X-ray
binaries. Such compact isolated accretors will probably have unusual
time-variability properties as their discs may be much larger than the typical
discs of X-ray binaries, and indeed they are missing the perturbing influence
of the secondary. On the other hand, accretion disc feeding in these sources
will be variable itself, leading to variability on variety of time
scales. Illumination of nearby gas clouds by X-rays may also lead to
reprocessed infra-red, optical and UV emission. Finally, the sources may leave
trails of denser and likely hotter gas behind them as they plough through the
gas.

\subsection{Quasars}

The right panel of Figure 1 shows that the emission from the cusp of stellar
mass remnants is bolometrically unimportant for a quasar accreting at a good
fraction of its Eddington limit. As already discussed, the result is not
surprising, as the total mass of the stellar mass black hole cluster is
$\approx 5$\% of the SMBH mass in this model.  However, the spectrum of
stellar mass black holes may be sufficiently different from that of the quasar
that the former may still contribute to some wavelength range. In particular,
the accretion disks of stellar mass black holes are systematically hotter than
those of much more massive quasar black holes \citep{Shakura73}. The black
hole cusp could then contribute to the soft X-ray band emission of quasars
where quasar disks are not expected to radiate much energy (except from a
corona or a jet).

\subsection{LLAGN}

Low Luminosity AGN are the systems where the accreting stellar mass black
holes may potentially complicate interpretation of the data, as the X-ray
luminosities of these systems roughly range from $L_X \sim 10^{39}$ erg/s to
$L_X \sim 10^{42}$. Further, unlike \sgra, emission of these black holes is
spatially unresolvable from that of the LLAGN. It is desirable to the broad
band spectrum of the black hole cusp in the future work. This may be non
trivial if some of the bright systems are in poorly understood super-Eddington
accretion rate regime, similar to the famous X-ray binary system SS443
\citep[e.g.,][]{Fabrika06}.

A way to distinguish between the SMBH and black hole cusp emission is via
time-dependent observations. The emission from a relativistic region of a SMBH
might vary on time scales of hours to days, as observed in both \sgra\
\citep{Baganoff01}, and in more distant AGN \citep[e.g.,][]{EdelsonEtal02}.
The emission of a large number of statistically independent black holes and
neutron stars should be considerably less variable.

\section{Conclusions}

The presence of stellar mass black holes and neutron stars in a cusp around a
SMBH is a robust theoretical prediction
\citep[e.g.,][]{Morris93,Miralda00,Freitag06}.  The underlying physics --
dynamical friction of these stellar remnants on a background stellar
population -- is very well understood. Here we attempted to estimate the
accretion rates with which these objects would be capturing gas from a
massive accretion disc. We also estimated the X-ray luminosities of these
objects. We found that the total luminosity of stellar mass remnant cusps
might be comparable to or larger than the luminosity of the central SMBH at
low gas densities. These conditions are plausibly met in LLAGN and the centre
of our Galaxy. Observational constraints on emission from these sources could
be used to constrain population of these sources in the central parsecs of
galaxies.

\end{document}